\documentclass[12pt,a4paper]{article}
\usepackage{tikz,amsmath,amssymb,extarrows,mathtools,graphicx,subfigure,setspace}
\usepackage[margin=2cm]{geometry}
\usepackage{cite}
\usepackage{setspace}

\usetikzlibrary{shapes,snakes}
\usetikzlibrary{decorations.markings}
\usetikzlibrary{decorations.pathreplacing}
\usepackage[colorlinks=true
,urlcolor=blue
,anchorcolor=blue
,citecolor=blue
,filecolor=blue
,linkcolor=blue
,menucolor=blue
,linktocpage=true
,pdfproducer=medialab
,pdfa=true
]{hyperref}

\newcommand{\bea}{\begin{eqnarray}}
\newcommand{\eea}{\end{eqnarray}}
\begin{document}
	\begin{center}
		
		{\Large {\bf On the Complexity of a Charged Quantum Oscillator} }
		\vspace*{.9cm}

			{Reza Pirmoradian$^1$ and
			Mohammad Reza Tanhayi$^{1,2,\star}$
		}
			\vspace*{.5cm}
		
			{$^1$ Department of Physics, Islamic Azad University  Central Tehran Branch, Tehran, Iran\\
			$^2$ School of Physics, Institute for Research in Fundamental Sciences (IPM), P.O. Box 19395-5531, Tehran, Iran}
		
		\vspace*{0.5cm}
		{E-mails: {\tt $^\star$mtanhayi@ipm.ir, rezapirmoradian@ipm.ir}}%
		\vspace*{1cm}
	\end{center}

\begin{abstract}
	
In this paper, we study the effect of both the electric and the magnetic fields on the rate of complexity growth. Our system is a charged quantum oscillator and over a period of time, we study the maximum dynamic evolution of quantum states which might lead to a strong bound on the rate of computation. We show that by turning on the electric field, the rate of complexity decreases whereas this rate has increasing behavior when the magnetic field is turned on. In this regard, we also find a critical value of the magnetic field, beyond which this rate changes its behavior drastically. 
\end{abstract}
\section{INTRODUCTION}

$\,\,\,\,\,$In the field of computing operations and information processing, two fundamental issues exist. The speed of processing information and the amount of memory capacity of a computer.  In other words, the problem of limits on computing speed and the amount of memory are challenging issues, and notably they might be characterized, respectively, by the energy and the number of degrees of freedom that the system can achieve. In this way, for a given quantum mechanical system, the limit on computing speed can be translated to the maximum number of distinct states that this system passes per unit time. Such states are supposed to be orthogonal; hence, the issue of computing the limitation of speed can be addressed using quantum mechanics. 

On the other hand, quantum mechanics teaches us to think that black holes are not `black' \cite{Hawking:1974rv}, but rather, if one considers a black hole as a data store or computational device, that compresses a given definite amount of energy, then that device can, indeed,
perform a certain number of  operations per second \cite{ref1}.  Furthermore, information processing has a theoretical upper limit,  more precisely, for a quantum system with a given average energy $E$, one can set a theoretical upper limit on the number of operations that can be performed in one second, which is given by $\frac{2E}{\pi\hbar}$. This limitation known as Lloyd's bound \cite{2000Natur.406.1047L}.

Quantum information theory is, indeed, a functional aspect of quantum mechanics and plays a crucial role in the development of the computer science. Interestingly enough, a concrete realization between black hole physics and quantum information theory has been argued. Namely, in the study of the physics of black holes, quantum information theory may play an important role.  For example, Bekenstein \cite{55} argued that black holes set a theoretical maximum on information storage, which means memory is bounded. Therefore,  understanding quantum information from a theoretical point of view would be an interesting issue or might even be a crucial one. In this way, entanglement entropy and complexity are important quantities (for more details, see Ref.s\cite{VanRaamsdonk:2013sza, Susskind:2014moa}). Interestingly, the black hole can, in principle, set fundamental limits on the density, entropy and computational complexity \cite{wer}. On the other hand, in a seminal work, Brown  {\it et al.} \cite{qqq} discovered a surprising connection between the action of the interior of the black hole and the above-mentioned upper limit of processing. In fact, what was argued goes as follows: the action of the interior of the black hole increases at a rate exactly equal to $\frac{2E}{\pi\hbar}$ and this led them to conclude that black holes produce complexity at the fastest possible rate. Thus, a relevant question to ask is the following: If the system is disturbed, can the system exceed this bound?

On the other hand in computer science, the computational complexity, or simply the complexity of an operation, has a simple definition:  it is the amount of resources required for running a problem, or it is the minimum of the complexities of all possible algorithms for this problem \cite{wert}. Moreover, quantum mechanics has a standard definition of complexity. In fact, complexity is a criterion that shows how difficult the task is. In principle, the complexity of a quantum state that
originates from the field of quantum computations is defined by the number of elementary unitary operations required to build up a desired state from a given reference state \cite{3}. An argument was also made that quantum complexity helped us to capture some certain features of the late-time behavior of eternal black hole geometries, as well \cite{4}.


 In Ref. \cite{Cottrell:2017ayj}, the authors consider the underlying Lloyd's bound
in relation to holographic complexity by making a distinction between orthogonalization and ‘simple’
gates and argue that these notions are useful for diagnosing holographic complexity. In fact, the amount of information that a physical system can store or process directly relates to the number of available system states.  Our main goal in this article is to examine Lloyd's bound for a quantum system under different conditions and then to compare the results with a system in which quantum states are not necessarily orthogonal. Given the fact that the quantum harmonic oscillator is one of the key issues in quantum mechanics, we intend to compute the variation of the complexity rate for this oscillator when the system is perturbed by electric, as well as, magnetic, fields. 

The layout of this paper is as follows: In Section 2, we briefly consider the orthogonality time for two simple uncoupled harmonic oscillators in the presence of magnetic and electric fields. In particular, we explore the minimum time needed for a given state to evolve into an orthogonal state. In Section 3, we look for any upper bound on the complexification. Finally, concluding remarks are given in Section 4.


\section{ORTHOGONALITY TIME IN A HARMONIC OSCILLATOR}
$\,\,\,\,\,$Let us consider two, simple, uncoupled harmonic oscillators subjected to both uniform electric and magnetic fields.\footnote{See also Ref. \cite{1} for related work. Note that throughout this paper,  we have used the natural units $\hbar=c=1$.} The corresponding Hamiltonian is then given by   
\bea
{H}=\frac{1}{2m}(p^2_x+p^2_y)+\frac{1}{2}m({w}^2+ \frac{q^2 B^{2}}{4m^2 c^2})(x^2+y^2)+\frac{qB}{2mc}(yp_x-xp_y)-\frac{q^2}{2 m\omega^2} \mathcal{E}^2,
\eea
where $m$ and $q$ stand for the mass and the charge of the system, respectively. The uniform electric and magnetic fields are supposed to be in the $x$ and the $z$ directions, respectively;  Also,  in writing the above Hamiltonian, we used the identity
$\vec{A}=\frac{B\hat{z}\times \vec{r}}{2}
$.
 The eigenvalue of a $2D$ charged harmonic oscillator is then given by
\bea
\mathbb{E}=\sqrt{{\omega}^2+ \omega_c^2} (n_1+n_2+1)- \omega_c(n_1-n_2)-2m\frac{ \omega_e^2}{\omega^2},
\eea
where $n_1$ and $n_2$ are positive integers or zero, also,  $\omega_e$ and $ \omega_c$ are defined by 
\bea
\frac{qB}{2m}\equiv \omega_c\,,\hspace{5mm}\frac{q\mathcal{E}}{2m}\equiv \omega_e.
\eea
\subsection{Minimum Rate of Orthogonality}
$\,\,\,\,\,$Now, let us compute the minimum time needed for any state of a given physical system to evolve into an orthogonal state. To do so, we take an arbitrary quantum state  as a superposition of energy eigenstates as follows: 
\bea
\mid\psi_0\rangle=\displaystyle\sum_{n_1=0}^{\infty}\displaystyle\sum_{n_2=0}^{\infty}c_{n_1,n_2}\mid n_1,n_2\rangle,
\eea
where we have assumed that the system has a discrete spectrum. Then, in a  straightforward manner, the time evolution of $\mid\psi_0\rangle$ can be written as follows: 
\bea
\mid\psi_{\tau_\bot}\rangle=e^{-iH\tau_{\bot}}\mid\psi_0\rangle=\displaystyle\sum_{n_1=0}^{\infty}\displaystyle\sum_{n_2=0}^{\infty}c_{n_1,n_2}e^{-i \,E_{n_1,n_2}\tau_{\bot}}\mid n_1,n_2\rangle.
\eea
To find the minimum time for orthogonality, let us define
\bea 
 S(\tau_{\bot})\equiv \langle\psi_{0}\mid\psi_{\tau_{\bot}}\rangle=\displaystyle\sum_{n_1=0}^{\infty}\displaystyle\sum_{n_2=0}^{\infty}{\mid c_{n_1,n_2}\mid}^2 e^{-iE_{n_1,n_2}{\tau_{\bot}}},\label{st}
\eea
where solving $S(\tau_{\bot})=0$ gives us the desired minimum time $\tau_{\bot}$. After doing some algebra and making use of the following relation for $x\geq0$
\begin{equation} 1-\frac{2}{\pi}[x+\sin(x)]\,\,\le \cos(x),
\end{equation} 
 one can deduce that
\bea\label{Tt}
1-\frac{2 \mathbf{E}\,\tau_{\bot}}{\pi}+\frac{2}{\pi}Im (S)\,\,\,\le Re(S),
\eea
where $\mathbf{E}$ is the average energy in the state $\mid\psi_0\rangle$. By imposing the orthogonality condition which leads to $Re S(\tau_{\bot})=Im S(\tau_{\bot})=0$, one finds that
the minimum time  $\tau_{\bot}$ required $\mid\psi_0\rangle$ to evolve into an orthogonal state is given by
\bea
\tau_{\bot}=\frac{\pi}{2 \mathbf{E}}.
\eea
This means our quantum system with energy $\mathbf{E}$ needs at least a time of $\tau_{\bot}$  to go from one state to an orthogonal state. This is, indeed, the Margolus-Levitin theorem \cite{aaaa}, which gives a fundamental limit on quantum computation. 

Now, the main aim is to compute the average energy in the state $\mid\psi_0\rangle$. In our case for a simple uncoupled  harmonic oscillator,  one has
\bea
\mathbf{E}&=&\langle\psi_0\mid H\mid\psi_0\rangle \cr \nonumber\\
&=&\displaystyle\sum_{n_1=0}^{N}\displaystyle\sum_{n_2=0}^{N-n_1}{\mid c_{n_1,n_2}\mid}^2\Big[\sqrt{{\omega}^2+ \omega_c^2}(n_1+n_2+1)- \omega_c(n_1-n_2)-2m\frac{ \omega_e^2}{\omega^2}\Big].
\eea
\begin{figure}\label{Fig1}
	\centering
	\includegraphics[scale=.28
	]{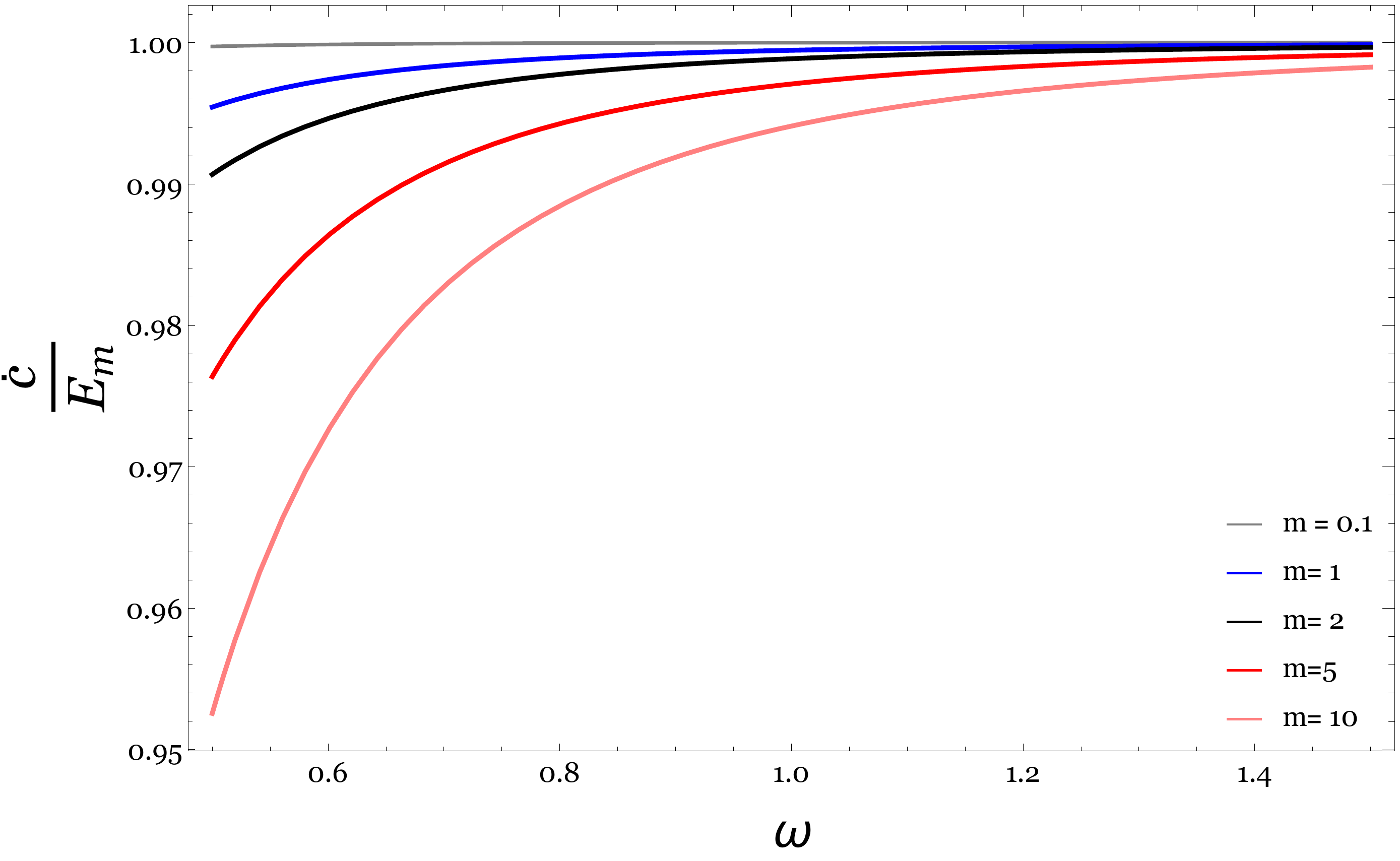}
	\includegraphics[scale=.28
	]{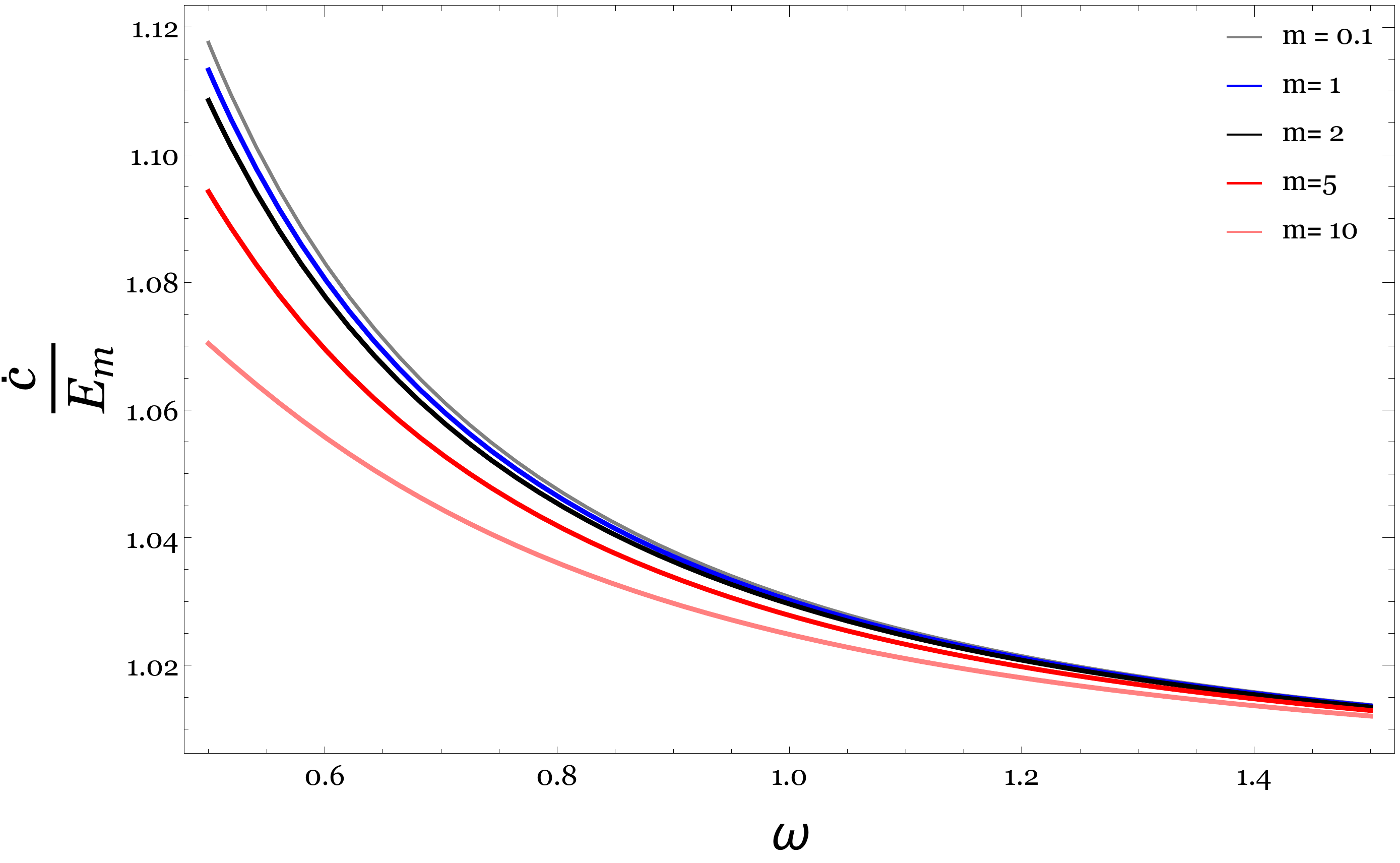}
	\caption{  Rate of normalized complexification of the quantum harmonic oscillator 
		for different values of mass for a fixed $N=200$. {\em Left panel}: $\omega_c=0.01$ and $\omega_e=0.2$. {\em Right panel}:  $\omega_c=0.25$ and $\omega_e=0.2$. }
\end{figure} 
To make calculations easier, one can start by
considering the set of evolutions that pass through an exact cycle of $N_t$ states that are supposed to be mutually orthogonal states at a constant rate  in time $\tau$; in this case, one has ${N_t}=\displaystyle\sum_{n=0}^{N}{(n+1)}$, where we have set $n=n_1+n_2$ and $ c_{n_1,n_2}=\sqrt{\frac{1}{N_t}}$. Therefore, one obtains
\bea
&&\mathbf{E}
=\displaystyle\sum_{n_1=0}^{N}\displaystyle\sum_{n_2=0}^{N-n_1}({\frac{1}{N_t}}) \sqrt{{\omega}^2+ \omega_c^2}(n_1+n_2+1)-\displaystyle\sum_{n_1=0}^{N}\displaystyle\sum_{n_2=0}^{N-n_1}({\frac{1}{N_t}})  \omega_c(n_1-n_2)-2m\frac{ \omega_e^2}{\omega^2}
\cr \nonumber\\
&&\hspace{.5cm}\displaystyle=\sum_{n=0}^{N}({\frac{n+1}{N_t}}) \sqrt{{\omega}^2+ \omega_c^2}(n+1)-2m\frac{ \omega_e^2}{\omega^2}\cr \nonumber\\
&&\hspace{.5cm}\displaystyle=(1+\frac{2N}{3})\sqrt{{\omega}^2+ \omega_c^2}-2m\frac{ \omega_e^2}{\omega^2}.
\eea
On the other hand, making use of the inequality in Eq.\eqref{Tt}, one can show that\footnote{Note in our consideration, we are dealing with large $N$ and also with an imposed electric field that satisfies the following limit, which, indeed, leads to non-negative energy:
	$$\omega_e\le\frac{\sqrt{3+2N}}{\sqrt{6m}}\omega^{3/2}.$$ }
\bea\label{T}
\tau_\bot\geq\frac{\pi}{2(1+\frac{2N}{3})\sqrt{{\omega}^2+ \omega_c^2}-4m\frac{ \omega_e^2}{\omega^2}}\,.
\eea 
This indeed sets a bound on the time required for  a given quantum system (charged harmonic oscillator subjected to both electric and magnetic fields) of energy $\mathbf{E}$ to go from one state to an orthogonal state. In what follows, we show that this bound potentially introduces an upper bound on the rate of complexity. 


\section{UPPER BOUND ON COMPLEXIFICATION}

$\,\,\,\,\,$In principle, the complexity of a quantum state
 is defined by the number of elementary unitary operations required to build up a desired state from a given reference state. In particular,  Lloyd in his seminal work \cite{2000Natur.406.1047L} showed that the speed with which a physical device can process information is limited by its energy and the amount of information that it can process is limited by the number of degrees of freedom it possesses. This limitation, which sets a fundamental upper
limit on the computation speed for a classical computer, is named as Lloyd's bound. Basically, the information process done by a computer is supposed to take a given
initial state to a final state via successive application of logic gates. In principle, a logical gate refers to an actual physical device that performs a logical operation. On the other, hand  any gate or operation takes some time $t$ to perform its task. If one implements a
Hamiltonian action $H_g$ to a task that is done by given gate, then the corresponding operation might be given by $U(t)=e^{iH_g\,t}$. Therefore, by definition, a unitary evolution takes the initial state $|0\rangle$ to the desired final state $U(t)|0\rangle$ after time $t$. Now, let us apply a sequence application of $n$ gates; namely, we have
\begin{equation}
|0\rangle\rightarrow U(t)|0\rangle.
\end{equation} 

In Ref.\cite{Cottrell:2017ayj}, the authors showed that for a series computation, one has
\begin{align}U(t)=\mathbf{T} \prod_i U(t_{i+1},t_i), \end{align}
where $U(t_{i+1},t_{i}) $ stand for an orthogonalizing gate and $\mathbf{T}$ is the time ordered operator. In the present case, the rate of complexification has a strong bound as follows:
\begin{align}\dot{\mathcal{C}}\leq \frac{1}{\tau_{\bot}}\label{cm},\end{align}
where  $\tau_{\bot}$ is the orthogonalization time of the system. Thus, in our case, the rate of complexification is given by 
\begin{align}
\dot{\mathcal{C}}\leq \Big(E_m\sqrt{1+\frac{\omega_c^2}{\omega^2}}-\frac{4m \omega_e^2}{ \pi \omega^2} \Big),
\end{align}
where we have defined $E_m=\frac{2}{\pi}(1+\frac{2N}{3})\omega.$ 

In figure.2, we have plotted the rate of complexity for different values of $\omega_c$ and $\omega_e$. As is shown from figure.1 and 2, the magnetic field has a critical value beyond which the rate of complexity changes its behavior drastically as
\begin{equation}\omega^{cri}_c=\frac{2\sqrt{3}}{(3+2N) \omega^2}\sqrt{m\omega_e^2[(3+2N) \omega^3+3m\omega_e^2]}\,\,.\label{criti}
\end{equation}
The results indicate that the normalized complexification of the quantum harmonic oscillator saturates to a definite value for large $\omega$. However, a critical value for magnetic field which defines the behavior of this saturation, exists. 

If one is interested in the case of a weak magnetic field near orthogonality of states, then the condition in Eq.\eqref{st} can be written as  $ S(\tau)\approx\epsilon e^{i\alpha}$, where $\alpha$ is a free parameter and $\epsilon$ is an infinitesimal. Therefore, in this case, one obtains
\bea 
\dot{\mathcal{C}}\lesssim \Big(E_m(1+\frac{\omega_c^2}{2\omega^2})-\frac{4m \omega_e^2}{ \pi \omega^2} \Big)\big[1-\epsilon\big(\frac{2}{\pi}\sin\alpha- \cos\alpha\big)\big];
\eea
on the other hand, a maximum value of $\frac{2}{\pi}\sin\alpha- \cos\alpha $ is $\sqrt{1+(\frac{2}{\pi})^2}$ is obtained, which leads to 
\bea 
\dot{\mathcal{C}}\lesssim \Big(E_m(1+\frac{\omega_c^2}{2\omega^2})-\frac{4m \omega_e^2}{ \pi \omega^2} \Big)\Big(1-\epsilon \sqrt{1+(\frac{2}{\pi})^2}\Big).
\eea
Figure.3 compares the rates of complexity for the orthogonal and the near orthogonal cases, where we have assumed that the magnetic field is small.

\begin{figure}\label{Fig2}
	\centering
	\includegraphics[scale=.38
	]{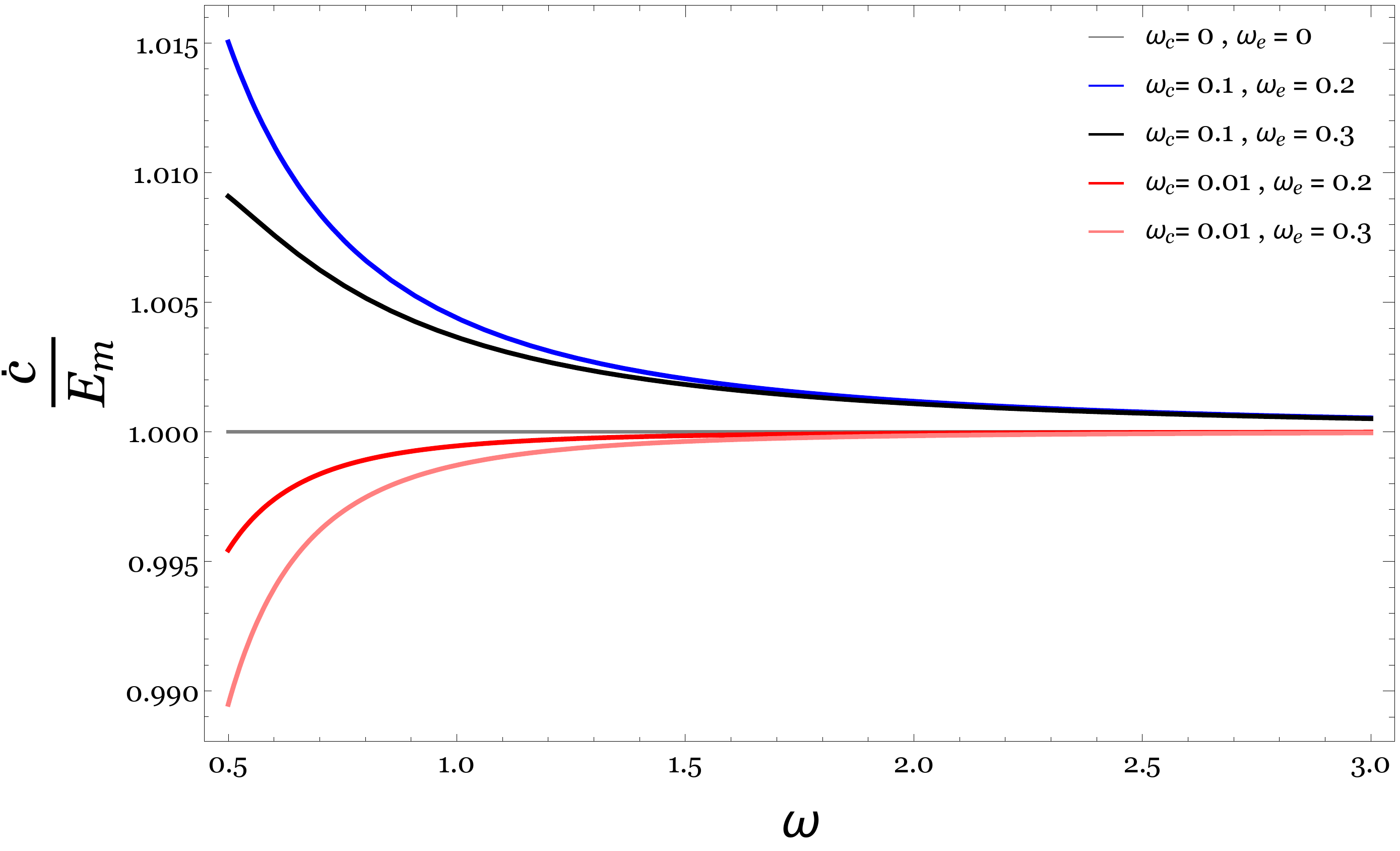}
	\caption{ Rate of normalized complexification of the quantum harmonic oscillator. When the magnetic field induction reached a definite critical value given by Eq.\eqref{criti}, the behavior of this rate changed. Note that we set  $N=200$ and $m=1$.} 
\end{figure}

\begin{figure}[htb!]\label{Fig3}
	\centering
	\includegraphics[scale=.4]{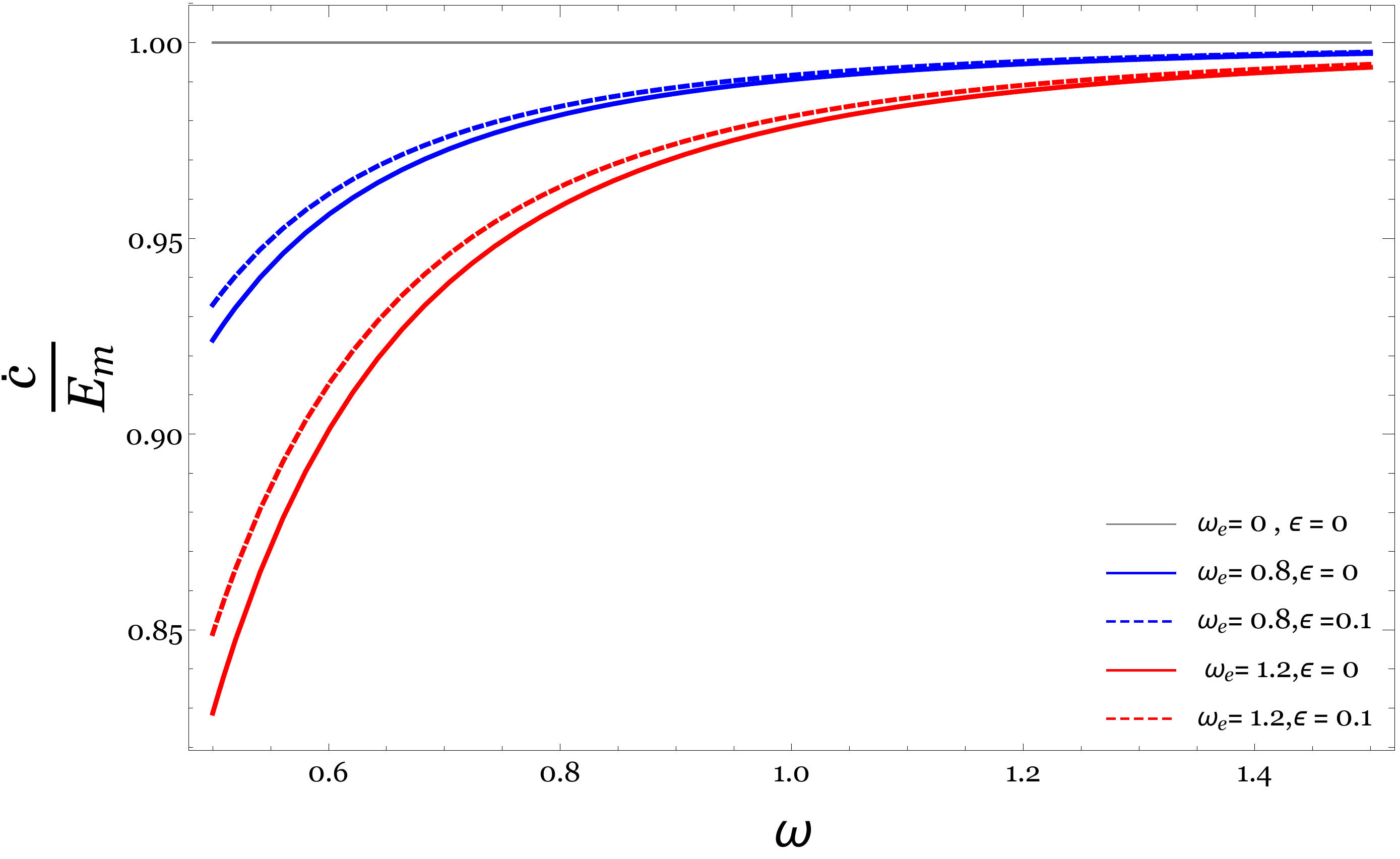}
	\caption{\label{figureone}Rate of variation in the complexity of the quantum harmonic oscillator for orthogonal (solid line) and near orthogonal (dashed line) states. Note that we set  $N=200$, $m=1$, $\omega_c=0$ (gray line) and $\omega_c=0.01$.}
\end{figure}


\section{REMARKS AND CONCLUSIONS}

$\,\,\,\,\,$Basically, the speed limit on information processing can be addressed by studying the bounds on the growth rate of the complexity, and in this paper, we discussed this issue for a charged harmonic quantum oscillator. We used the minimal time of orthogonalization similar to what used in the Margolus-Levitin
bound, where $\langle E \rangle$, the expectation value of the energy of the system, plays a crucial role. On the hand, for Lloyd's bound (the bound on the rate of computation), the minimal time to perform a task is controlled by the energy $E=\langle H \rangle$, which defines a limitation on the computations. Actually, the computation speed might be given by the number of operations per unit time step or the time rate of change of the complexity. In this way, Lloyd's bound defines the upper bound as $\frac{d}{dt}{\cal
	C}\le \frac{2E}{\pi\hbar} $.
The computational speed is bounded from above by the instantaneous energy in a unit time step. This potentially gives a bound on the growth rate of the holographic complexity, as is discussed in the context of AdS/CFT. In this notation all operations (or equivalently all gates) can be implemented quantum mechanically, so inputs and outputs which are classical states, can occur at any step. These states are orthogonal and have no superpositions with each other.

The main goal of this paper was to investigate the  complexity of a charged harmonic oscillator by orthogonalizing states. This was achieved by studying the maximum speed of dynamical evolution, followed by finding the maximum number of distinct states that the system could pass through per unit of time. For a quantum system, the distinct states can be understood by using the orthogonality of states. We considered the harmonic oscillator in the presence of both magnetic and electric fields and found the minimum time needed for any state of our system to evolve into an orthogonal state.  We observed that the time of orthogonality for small and large magnetic fields behaved differently and that the rate of complexity was increased/decreased by turning on the magnetic/electric field.

Recently, the study of the complexity of charged black holes from the holographic point of view and the investigation this issue using quantum field theory have received much attention (see for example  Ref.s\cite{qqq,Brown:2015bva,Doroudiani:2019llj}). In this way considering the rate of change of the complexity of a charged quantum harmonic oscillator in the presence of external fields seems to be important.
Motivated by this fact, as a future work, we
will implement this model for computing the rate of complexity  for  a  charged  thermofield  double  state of  free
real scalar quantum field theory in the presence of a background electric field to investigate the effect of such fields on complexification.

\section*{ACKNOWLEDGMENTS}

$\,\,\,\,\,$We would like to kindly thank M. Alishahiha and A. Naseh for useful comments and discussions on related topics. We also acknowledge  A. Akhavan and  F. Omidi for some comments. We also thank the referee for his/her very useful comments.


\end{document}